\documentstyle[12pt]{article}
\newcommand{\be}{ \begin{eqnarray}}
\newcommand{\ee}{\end{eqnarray}}
\newcommand{\beno}{ \begin{eqnarray*}}
\newcommand{ \eeno}{\end{eqnarray*}}
\newcommand{\raf}[1]{(\ref{#1})}
\newcommand{\lam}{\Lambda(1405)}
\begin{document}
\bibliographystyle{try}
 \begin{titlepage}
\hspace{11cm}
{\large SUNY-NTG 94-27}
\vspace{.7cm}
 \begin{center}
\ \\
{\large {\bf K$^-$-proton scattering and the $\Lambda (1405)$ in dense matter}}
\vspace{2cm}
\ \\
{\large Volker Koch }
\ \\
\ \\
{\it Physics Department, State University of New York\\
Stony Brook, NY 11794, U.S.A.}\\
\ \\
\ \\

\vspace{2cm}
{\large {\bf Abstract}}\\
\vspace{0.2cm}
\begin{quotation}
The scattering of antikaons with nucleons is studied in the nuclear
environment. Describing the $\lam$ as a $K^-$ proton bound state, we find,
that due to the Pauli blocking of intermediate
states, the mass of the $\Lambda (1405)$ is shifted upwards in energy,
above the $K^-$ proton threshold and and its width is somewhat
broadened. As a
consequence the s-wave $K^-$--nucleon scattering length turns attractive at
finite nucleon density ($\rho \geq 0.25 \rho_0$) leading to a mean field
potential for the $K^-$ of about $\sim -100 \, \rm MeV$ in nuclear matter at
ground state density. Consequences for Heavy Ion collisions and possible
experimental checks for the structure of the $\lam$ are discussed
 \end{quotation}
\end{center}
\end{titlepage}
\newpage
\section{Introduction}
Recently the physics of kaon-nucleon scattering has received considerable
interest because of possible implications for the structure of neutron stars
and
relativistic heavy ion collisions. First speculations about the
possibility of kaon condensation have been put forward by Kaplan and Nelson
\cite{NK87} and Politzer and Wise \cite{PW91}
on the basis of a simple mean field evaluation
of a Lagrangian derived from chiral perturbation theory. At mean field
level this Lagrangian predicts two potentials for the kaons.
(i) A scalar term due to explicit
chiral symmetry breaking, which is attractive for both $K^+$ and $K^-$,
and which involves the kaon-nucleon-sigma term. (ii)
A vector term due to vector
meson $(\rho,\, \omega)$ exchange, which is attractive for the $K^-$ and
repulsive for the $K^+$. For the $K^+$ these two terms essentially
cancel each other leading to a small repulsive potential for the $K^+$.
For the $K^-$, on the other hand, scalar and
vector term add up to a large attractive mean field potential which, depending
on the choice of the sigma-term, leads to a critical density of $2-5 \,
\rho_0$ for kaon condensation, where $\rho_0$ is the nuclear matter ground
state
density. Based on this model for the $K^-$ mean field Brown et al.
\cite{BLR94}
found that the nuclear equation of state softens considerably
resulting in qualitatively new properties of neutron stars and leading
to the speculation of Brown and Bethe \cite{BB94} about the existence
of small mass black holes. Rho and collaborators \cite{LJM94},
furthermore, have
carried the work of Kaplan and Nelson to higher order in the chiral counting
but
did not find qualitatively new features. In heavy ion physics, the mean fields
would help to explain the observed slope difference between $K^+$ and $K^-$
spectra at AGS energies ($14 \, \rm GeV/u$) \cite{FKB93}
as well as the production of subthreshold $K^+$ at SIS energies ($ \sim 1 \,
\rm GeV/u$) \cite{FKL94}.

There is, however, a phenomenological problem with the approach of Kaplan
and Nelson. It is well known that s-wave the $K^-$ nucleon scattering
length is repulsive at threshold \cite{Mar81} $(a_{K^-N} = -.15 \, \rm fm)$.
Therefore, at very low densities, where the impulse approximation should be
valid, we would expect a repulsive mean field of the form $ U \sim -4 \pi
a_{K^-N} \rho $.
Kaplan and Nelson on the other hand would predict an attractive
$K^-$
nucleus optical potential even at arbitrarily
low density, which is equivalent to
say, that their scattering amplitude obtained at tree level does not agree with
experiment. The reason for this disagreement is the existence of the
s-wave, isospin $I=0$ $\lam$
resonance in the $K^-proton$-channel just
below threshold.
Scattering through this resonance, which
can be understood as a $K^-p$ bound state in the continuum
\cite{DWR67,SW88}, gives rise to a repulsive contribution to
the scattering amplitude at
threshold. This phenomenon is well known in nuclear physics: the p-n scattering
length in the deuteron channel ($I=0,S=1$) is repulsive due to the existence of
the deuteron as a bound state\footnote{Actually simple arguments from
low energy scattering show, that the existence of a bound state below threshold
always leads to a repulsive scattering length \cite{PB74}.}.

The mean field potential of Kaplan and Nelson simply ignores the existence of
the $\lam$ resonance and, therefore, gives the wrong low density behavior
for the s-wave $K^-$-nucleus optical potential. The question then, of course,
is
to which extent this model can be trusted at higher densities.
In a recent work Lee et al. \cite{LJM94}
 have included an explicit resonance term for the
$\lam$ in their calculation of the K-N scattering amplitude in chiral
perturbation theory. After adjusting the counter terms, they conclude that
the existence of the $\lam$ does not affect the conclusions about the
existence of the $K^-$-condensate, because the kaon-wave in the condensate
is very much off shell, below the $\lam$ resonance.
While this reasoning may be in principle correct, it of course relies heavily
on
the off shell properties of the $K^- -p$ amplitude. The use of mean field
potentials, of course, would be justified much better, if the $\lam$ would not
exist at all in dense matter. This is the case in our example of the deuteron.
There, we know, that nuclei are not made out of deuterons, and that the mean
field approximation works very well.
The reason is that the Pauli principle
destroys the deuteron in the nuclear environment.
Proton and neutron cannot form
a bound state, because the momentum space their wave function
would occupy is already
occupied by other nucleons in the environment. In more technical terms, the
intermediate states of the p-n t-matrix are Pauli blocked in matter such that
no pole below threshold is formed.
Consequently, if the $\lam$ can be understood as a $K^--p$ bound state, similar
effects due to Pauli blocking should change its properties in the nuclear
medium.

It is the purpose of this article to quantitatively investigate, how the
nuclear
environment changes the properties of the $\lam$ and what the consequences for
the in medium properties of the $K^-$ are.
This investigation naturally relies
on the assumption that the $\lam$ is indeed a $K^--p$ bound state and not a
genuine 3-quark state.  Therefore, all changes we predict due to the effect of
Pauli blocking of the proton in nuclear medium are absent if the $\lam$ is a
genuine
3-quark state. Consequently, the medium effects discussed here, can be used
to {\em experimentally} distinguish between the two pictures for the $\lam$. As
we will show below, for densities $\rho \leq \rho_0$ the effect of the medium
it to move the position of the $\lam$ up in energy, while its width remains
essentially constant. This behavior can be observed in experiment by
photo production of the $\lam$ in a nucleus via the reaction
\be
\gamma + p \rightarrow K^+ + \lam
\label{eq.1.1}
\ee
We estimate, that this experiment should be feasible at CEBAF energies.

This article is organized as follows. In the next section we  will briefly
discuss the phenomenology of the $K^-p$ scattering and introduce the model
which we will use to describe the $\lam$.
Section 3 will be devoted to the consequences for the $K^-$ self energy.
We
will briefly address the issue of finite temperature, which is relevant to
the phenomenology of relativistic heavy ion collisions. Finally, in section 4
we will discuss
possible experimental checks for the structure of the $\lam$
and implications of our results for the
physics of heavy ion collisions as well as for astrophysics.

\section{The model}
The idea to understand the $\lam$ as a $K^-p$ - bound state is not at all new.
Already back in the sixties \cite{DWR67} attempts have been made to describe it
as a bound state in a Yukawa potential due to vector meson exchange. More
recently Siegel and Weise investigated the $K^-$-nucleon scattering
by using a separable potential model as well as a local Yukawa interaction.
Although some details of the amplitude needed additional ingredients, such as
explicit SU(3) breaking as well as a weakly coupling s-channel resonance
slightly below threshold, the cross features of the $I=0$ amplitude and the
position of the $\lam$ resonance could be reproduced well with the separable
potential alone. Finally, a very detailed calculation in the spirit of the
Bonn-meson exchange model has been carried out by the J\"ulich group
\cite{MHS90}.

In this work, we want to restrict ourselves to a separable potential model
for the $K^-p$ interaction, because it can be solved algebraically in
momentum space and the Pauli blocking can be incorporated
easily. It also makes the effect of the Pauli blocking more transparent and
certainly will provide the correct qualitative picture and a reasonable
quantitative estimate of the effects to be expected from a more detailed
calculation using e.g. the J\"ulich approach. Since the $K^-N$-system strongly
couples to other channels, such as $\Sigma \pi$, we have to treat the  full
coupled channels problem.

Siegel and Weise have demonstrated that around threshold a nonrelativistic
description of the $K^-$-nucleon system works well. We, therefore, will
restrict ourselves to a nonrelativistic treatment of the problem and, hence,
have to solve the following Schr\"odinger -- type equation.

\be
\nabla^2 \psi_i(r) + k_i^2 \psi_i(r) - 2 \mu_i \int V_{i,j}(r,r') \, \psi_j(r')
d^3r' = 0
\label{eq.2.1}
\ee
where $\psi_i(r)$ represents the wave function and $\mu_i$ the reduced mass for
the channel $i$.
The center of mass momenta $k_i$ of channel $i$ are related to the total
energy $E$ by
\be
k_i^2 = \frac{[E^2 - (M_i - m_i)^2] [ E^2 - (M_i + m_i)^2]}{4 E^2}
\label{eq.2.1b}
\ee
where $m_i$ and $M_i$ are the masses of meson and baryon in channel $i$.
Since we are interested in the isospin $I=0$ channel, only the
$K^- p$ and the $\Sigma \pi$--channels contribute.

For the separable potential we use the following ansatz in momentum space
\be
V_{i,j}(k,k') &=& g^2 C_{i,j} \,\, v_i(k) \, v_j(k')
\nonumber \\
              &=& \frac{g^2}{\Lambda^2} C_{i,j} \Theta(\Lambda^2 - k^2 )
                  \Theta(\Lambda^2 - k'^2 )
\label{eq.2.2}
\ee
Contrary to ref \cite{SW88}, in which  a Yukawa form for $v_i(k)$ was used,
we use
a sharp cutoff in momentum space. As we will show, this potential is sufficient
to reproduce the $K^-p$ amplitude in the region of interest and we feel that
the medium effects we are about to study become more transparent, when using
this simpler potential.

The $KN$ scattering amplitude is best studied by solving for the so called
T-matrix, which is equivalent to solving the Schr\"odinger equation for the
scattering problem (see e.g. ref. \cite{EK80}). Following \cite{EK80}, the
T-matrix is given by
\be
T_{i,j}(k,k',E) = V_{i,j}(k,k') + \sum_l\,
                 \int \frac{d^3q}{(2 \pi)^3} V_{i,l}(k,q)
                \frac{1}{E - m_l - M_l - q^2/(2 \mu_l)} T_{l,j}(q,k')
\label{eq.2.3}
\ee
where the energy $E$  is related to the eigenvalue $k_i^2$ of the
Schr\"odinger equation \raf{eq.2.1} given by eq. \raf{eq.2.1b}.
In case of a separable potential this integral equation is readily solved by
\be
T_{i,j}(k,k',E) = g^2 \,v_i(k) \, v_j(k')
\left[ (1- C \cdot G(E))^{-1} \cdot C \right]_{i,j}
\label{eq.2.5}
\ee
where we have defined the `propagator' matrix
\be
G_{i,j} &=& diag(g_i);
\nonumber \\
g_i(E) &=&  g^2 \,
          \int \frac{d^3 p}{(2 \pi)^3} \frac{v_i^2(p)}{E - m_i - M_i  -
p^2/(2 \mu_i)}
\nonumber \\
&=&\frac{1}{2 \pi^2} \frac{g^2}{\Lambda^2}
 \int_0^\Lambda \frac{p^2 \, dp}{E - m_i - M_i
- p^2/(2 \mu_i)}
\label{eq.2.6}
\ee
Here we have inserted our choice for the potential $v(k) =
\frac{\Theta(k^2-\Lambda^2)}{\Lambda}$.
The scattering amplitude is directly related
to the T-matrix by
\cite{EK80}
\be
f_{i,j}(k,k') = - \frac{\mu_i}{2 \pi} T_{i,j}(k,k')
\label{eq.2.7}
\ee

Finally, for the interaction matrix $C_{ij}$, we use the standard result
derived from $SU(3)$ flavor symmetry (see e.g. ref \cite{DWR67,SW88})

\be
C_{i,j} =
\begin{array}{cc}
\begin{array}{cc}
K^- p & \pi \Sigma
\end{array} & \\
\left(
\begin{array}{cc}
-\frac{3}{2} & -\frac{\sqrt{6}}{4}\\
\\
-\frac{\sqrt{6}}{4} & -2
\end{array}
\right)
&
\begin{array}{c}
K^-p\\
\\
\pi \Sigma
\end{array}
\end{array}
\label{eq.2.8}
\ee
where it is assumed that all interactions are mediated by vector meson
exchange.

In fig. \ref{fig.2.1} we compare the results, using three different cutoffs,
for
the $I=0$ $K^-N$ forward scattering amplitude $f(\omega) = f(k=k',\omega)$ with
that extracted from experimental data by Martin \cite{Mar81}. The agreement
with the `data' is reasonably good and essentially independent from the choice
of the cutoff. The coupling $g$, of course, depends on the cutoff chosen and
the values needed to reproduce the data are listed in table \ref{tab.2.1}.

\begin{table}
\begin{center}
\begin{tabular}{l|l}
$\Lambda \, [\rm GeV ]$ & $g^2 /4 \pi$ \\ \hline
1.0  & 1.733 \\
0.78 & 1.425 \\
0.6  & 1.2
\end{tabular}
\end{center}
\caption{Combinations of cutoff $\Lambda$ and coupling $g$ leading to the
amplitudes shown in fig. \protect\ref{fig.2.1}.}
\label{tab.2.1}
\end{table}

Having demonstrated that our model,
independent from the cutoff, can reproduce the data in free space we now can
proceed to finite density.

\section{Finite density}
\label{sec.3}
In the nuclear medium, because of the Pauli blocking, the intermediate
proton states with momenta $p \leq k_f$, where $k_f$ is the fermi momentum, are
forbidden. As a consequence the propagator for the $K^-p$ intermediate state
$g_1(E)$ \raf{eq.2.6}
 changes to
\be
g_1(E,k_f) = \frac{1}{2 \pi^2} \frac{g^2}{\Lambda^2} \int_{k_f}^\Lambda d p \,
p^2 \frac{1}{E - m_{Kaon} - m_{proton} - p^2/(2 \mu_{Kp})}
\label{eq.3.1}
\ee
while that for the $\pi \Sigma$ remains unaffected by the nuclear environment.
Of course nucleon and kaon are also subject to mean field forces,
which changes their
self energy in the medium. In a more consistent treatment along the line of a
Brueckner Hartree-Fock scheme, both effects need to be taken into account
together. This, however, is beyond the scope of this article and will be
addressed in a more complete investigation.

In order to qualitatively see the effect of Pauli blocking of the intermediate
states, let us ignore the coupling to the $\pi \Sigma$--channel for a moment.
In this case from eq. \raf{eq.2.6} the binding energy of the $\lam$ is given by
the solution of the integral equation
\be
1 - \frac{1}{2 \pi^2} \frac{g^2}{\Lambda^2} \int_{0}^\Lambda d p \,
p^2 \frac{\Theta(k_f - p)}{E - m_{Kaon} - m_{proton} - p^2/(2 \mu_{Kp})}
= 0
\label{eq.3.2}
\ee

The eigenvalue $E$ obtained from this equation is plotted as a
function of the nuclear density $\rho = \frac{2}{3 \pi^2} k_f^3$ in fig.
\ref{fig.3.1} for two different cutoffs $\Lambda$.
Here the couplings $g$ have been readjusted such that the empirical value
at $\rho = 0$ of $E = 1.405 \, \rm GeV$ is obtained. This readjustment
is necessary since we have ignored the $\pi \Sigma$--channel.
We see that the position of the $\lam$ moves up in energy with increasing
density and reaches the $K^-p$--threshold $(\omega = 0)$ at $\rho \simeq 0.5
\rho_0$.
This behavior can be easily understood from eq. \raf{eq.3.2}. The support of
the integral is reduced with increasing density reflecting the truncation of
momentum space due to Pauli blocking. In order to still satisfy eq.
\raf{eq.3.2} the denominator in the integrand has to be reduced by increasing
the energy eigenvalue $E$.
Notice, that even if the energy eigenvalue of the $\lam$ turns
positive, the $\lam$ cannot decay into a $K^-$ and proton, since the final
state
for the proton is Pauli-blocked.
As can be seen from eq. \raf{eq.3.2}, the integral acquires an imaginary part
only if $\omega \ge k_f^2/(2 \mu)$. This, however, does not happen for the
densities considered here $(\rho \leq 2 \rho_0)$.

While the decay of the $\lam$ is inhibited by the nuclear medium even
above threshold, the shift of its energy/mass changes the real part of the $K^-
p$--scattering amplitude. Once the $\lam$ has moved above threshold, the real
part changes sign giving rise to an attractive potential for a low energy
$K^-$ in nuclear matter.
Finally notice that the cutoff dependence of the results again is very small.

In this simple one channel model it is easy to include effects of finite
temperature, which are relevant for the physics of relativistic heavy ion
collisions. This is done by replacing the Theta-function $\Theta(k_f-p)$
with the fermi distribution function in the integral of eq. \raf{eq.3.2}. The
resulting eigenvalues are plotted in fig. \ref{fig.3.1} (dashed--dotted line)
for a temperature of $T = 150 \, \rm MeV$. We find that the shift of the mass
is smaller than in the zero temperature case, simply because at finite
temperature the Pauli-blocking of the low momentum states is less efficient.
Furthermore, once the $\lam$ has reached the $K^-N$ threshold, its decay is not
entirely Pauli-blocked and, hence,  the eigenvalue picks up an imaginary
part. This is the reason why we have plotted the dashed-dotted curve only up to
densities of $\rho =\simeq 1.3 \rho_0$. Beyond that value, the eigenvalue is
complex.
Although the shift in mass at finite temperature is reduced it is still
sufficient to warrant an attractive optical potential for a s-wave $K^-$ in a
fireball of this temperature.

In order to study the density dependence of the $K^-$ optical potential more
quantitatively, we have to evaluate the full $K^-p$ T-matrix, including the
coupling to the $\pi \Sigma$, channel in the nuclear medium. This is done by
evaluating the T-matrix according to eq. \raf{eq.2.5} with $g_1(E,k_f)$
given by eq. \raf{eq.3.1}.
In fig. \ref{fig.3.2}
we show the real and imaginary part of the $I=0$ $K^-p$ scattering
amplitude as a function of the energy for the free case (full line) as well as
for densities $\rho = 0.5 \rho_0$, $\rho = \rho_0$, and $\rho = 2 \rho_0$.
A cutoff of $\Lambda = 1 \, \rm GeV$ has been used. The
upward shift of the $\lam$ is clearly visible in the imaginary part of the
amplitude and the change in sign of the real part at threshold is nicely
demonstrated. Notice, that the width of $\lam$ changes very little, because of
the aforementioned blocking of the $K^-p$ decay channel. Also the relative
increase in the $\pi \Sigma$ phase space due to the increase in mass of the
$\lam$ is small ($\frac{\Delta p}{ p} \simeq 0.2)$. The long--dashed curve
corresponds to the result obtained for $\rho = 0.5 \rho_0$ with a smaller
cutoff of $\Lambda = 0.78 \, \rm GeV$. Again, the cutoff dependence is small.

In order to determine the $K^-$ optical potential in nuclear matter, we have to
include the isospin $I=1$ contribution as well
\be
U_{tot} = \frac14 U_{I=0} + \frac34 U_{I=1}
\label{eq.3.3}
\ee
Since in the energy region of interest there are no resonances which could be
affected by the nuclear medium similarly to the $\lam$, we determine the $I=1$
contribution to the optical potential by the lowest order impulse approximation
\be
2 m_k \, U_{I=1}(\rho) = - 4 \pi (1 + \frac{m_k}{m_N})\, a_{I=1} \, \rho
\label{eq.3.4}
\ee
where $a_{I=1} = 0.37 + i \, 0.6 \, \rm fm$
is the empirical scattering length for
the isospin $I=1$ channel.
The $I=0$ contribution we determine by integrating over the fermi sphere of the
nucleons
\be
U_{I=0} = -8 \pi (1 + \frac{m_k}{m_N}) \,
 \int d^3k \, \Theta(k_f-k) f(\omega = \sqrt{(m_p+m_k)^2 + \frac{k^2
m_k}{m_p}}\,\,)
\label{eq.3.5}
\ee
taking into account some of the fermi momentum corrections, which we find to be
small.

In fig. \ref{fig.3.3}, we have plotted the resulting optical potential as a
function of density. We find that the real part turns attractive for densities
$\rho \ge 0.25 \rho_0$.
Also the imaginary part increases with density, but it is still comparable in
magnitude to the real part, allowing for a quasiparticle approximation of the
kaon. Notice, that our result for the optical potential at $\rho = \rho_0$ of
$V \simeq -100 - i\, 100 \, \rm$ is right between the two different fits
from the analysis of kaonic atoms by Friedman
et al. \cite{FGB93}. They give a value of $V \simeq -73 -  i\, 109
\, \rm MeV$ for their effective `$t\rho$'--fit and $V \simeq -188 -  i \,73
\, \rm MeV$ for their density dependent potential fit.

It has has  been shown in
ref. \cite{BWT78} and more recently \cite{MHT94}, that an attractive real part
of the optical potential can be achieved
microscopically by requiring that the $\lam$ is subject to an attractive mean
field of $\sim -10 \, \rm MeV$ at
nuclear matter density, which is to be compared
with a standard value of $\sim -60 \, \rm MeV$ for the nucleon mean field.
Consequently, the $\lam$
is shifted upwards by $\sim 50 \, \rm MeV$ with respect to the in medium $K^-N$
threshold, and thus provides the necessary attraction, as it is now above
the in medium threshold. By looking at fig. \ref{fig.3.2} we see that our
calculation also predicts an upward shift of the $\lam$ of about $\sim 50 \,
\rm MeV$ with respect to the $K^-N$ threshold, in nice agreement with the
phenomenological analysis\footnote{In our calculation we have not taken the
mean field for the nucleon into account. Inclusion of mean field
potential for the nucleons would essentially shift the nucleon and the $\lam$
by the same amount, leaving, however, their mass difference unchanged.}.

On the other hand, assuming that the $\lam$ it is a genuine three quark
state, simple counting of the light quarks  would suggest that the
mean field potential for the $\lam$, is about $2/3$ of that of the nucleon,
leading to a relative shift of $\sim 20 \, \rm MeV$ of the $\lam$.
The analysis of the Kaonic atoms, therefore, may be the first
phenomenological hint, that the $\lam$ should indeed be viewed as a $K^- p$
bound
state. In the following section we will discuss a more direct way to measure
the shift of the $\lam$ in matter and, thus, to experimentally investigate the
nature of the $\lam$ state.

\section{Possible experimental test}

The position of the $\lam$ in matter actually can be measured via the reaction

\be
p+\gamma \rightarrow K^+ \lam
\ee

The missing mass spectrum of the $K^+$ then has a peak at the mass of the
$\lam$. These measurements have already been carried out back in the seventies
on a hydrogen target \cite{BDE71}.

Here, we propose to carry out these measurements using nuclear targets in order
to determine a possible shift of the mass of the $\lam$ in matter. These
measurements require a photon beam (tagged photons) of energy $E_{\gamma}
\geq 2
\, \rm GeV$ and, therefore, can be carried out at CEBAF.

If a shift of the $\lam$ of $\Delta M \geq 30 \, \rm MeV$ can be established,
it would have at least two interesting implications:
\begin{itemize}
\item Since the $\lam$ would then be above the $K^-N$ threshold, an
s-wave $K^-$
will feel an {\em attractive} optical potential in matter by scattering through
a intermediate $\lam$-state. This would be an experimental support for
the conjectures about kaon condensation in dense matter \cite{NK87,PW91}.
It furthermore would confirm the microscopic model for the $K^-$ optical
potential needed to fit the data from kaonic atoms \cite{BWT78,MHT94}.
While the spectroscopy of kaonic atoms measures directly the $K^-$ optical
potential, it is mostly sensitive to the low density region
in the surface of the
nucleus. The experiment discussed here, on the other hand,
probes the interior of the
nucleus, and thus nuclear matter density, provided the target nucleus is large.

Notice, that these implications are independent of the nature of the $\lam$, in
particular they
are independent of the model for the $\lam$ used in this article.
\item As far as the nature of the $\lam$ is concerned
an observed large shift ($\sim 50 \, \rm MeV$) in its mass
would help to answer this question. As already briefly discussed in section
\ref{sec.3}, from simple quark counting we expect that a genuine 3-quark state
will be shifted by only $\sim 20 \, \rm MeV$.
Contrary to that, the $K^-N$ bound
state, which is subject to Pauli-blocking of the intermediate state, will be
shifted by $ \sim 50 \, \rm MeV$, as discussed in the previous section.
This argument  can actually be sharpened somewhat. A reasonable estimate on the
relative shift of a 3-quark-$\lam$ should be that of the lowest $\Lambda$
state. From the analysis of the binding energies of $\Lambda$-hypernuclei the
depth of the mean field potential for the $\Lambda$ has been extracted to be
$V_\Lambda = - 28 \, \rm MeV$ \cite{MDG88}
implying an upward shift of $\sim 20-30 \, \rm
MeV$ with respect to the nucleon.
Furthermore, the same measurement as described here, can also be done for the
$\Lambda$ and have already been proposed in the literature (see e.g.
\cite{Dov92}.
On the theoretical side, the prediction of the shift of the $K^-N$ can be
improved by carrying out a selfconsistent Brueckner-Hartree-Fock type
calculation based on a more realistic model for the $K^-N$
interaction.
\end{itemize}

Using a nuclear target the signal may be distorted through fermi momentum
corrections, rescattering of the $K^+$, the cross section of which, however, is
small, and because of the nuclear surface, where the density and, thus, the
shift of the $\lam$ is small. To estimate these corrections, we have performed
a Monte Carlo calculation, which takes all these effects into account. We have
assumed that the matrix element for the process
$\gamma + p \rightarrow K^+ \lam$ is independent of energy and  isotropic
in the
c.m. frame. For the mass distribution of the $\lam$ we have chosen a
standard Breit Wigner Form. For the rescattering of the $K^+$ in the nuclear
medium we have assumed a cross section of $ 10 \,\rm mb$.
The resulting invariant mass distribution is shown in fig. \ref{fig.4.1}. The
full histogram shows the spectrum assuming no changes in mass and in the width
of the $\lam$ while the dashed histogram shows the resulting spectrum assuming
as mass shift of $30 \, \rm MeV$ and an increase of the width by $10 \, \rm
MeV$. The shift in the spectrum is still clearly visible and we conclude
that in spite of the corrections discussed above a mass shift of $\Delta M \geq
30 \, \rm MeV$ can be observed in this type of experiment\footnote{Notice, that
this experiment is sensitive to the mass {\em difference}
 between the $\lam$ and the
proton and not to the absolute position of the $\lam$. Therefore, an overall
shift of the same magnitude for both, the nucleon and the $\lam$,
due to the nuclear mean field, which
we have not taken into account in our calculation, does not change the outcome
of the experiment. Consequently the effect of the Pauli-principle and its
implications for the nature of the $\lam$, can be observed in experiment,
provided that the {\em relative} position of the $\lam$ with respect to the
nucleon has changed in the medium compared to the free case.}.

\section{Conclusions}
In this article we have investigated the properties of the $\lam$ resonance in
the nuclear medium and its consequences for s-wave $K^-$-nucleon scattering.
Assuming that the $\lam$ is a $K^-p$ bound state, we could show that its mass
increases with density as a result of the Pauli blocking of its wavefunctions
As a consequence, the $K^-N$ s-wave scattering amplitude turns attractive at
densities $\rho \geq 0.25 \rho_0$ in agreement with the analysis of kaonic
atoms. Our findings also support conjectures concerning a possible $K^-$
condensation in dense matter.

The resulting values for the $K^-$ optical potential are of comparable
size with
the numbers extracted from a fit to kaonic atoms by Friedman et al.
\cite{FGB93}. The shift of the $\lam$-resonance with respect to the
$KN$--threshold is in agreement with that needed in microscopic models for the
optical potential of kaonic atoms \cite{BWT78,MHT94}. In particular, our model
provides a dynamical explanation for this shift, which is based on the Pauli
blocking of the $K^-p$ wavefunction and, thus, on the compound nature of the
$\lam$.

We finally have proposed an experimental check for the shift of the $\lam$,
which should be feasible at CEBAF. If a shift larger that $\sim 30 \, \rm MeV$
could be established, an independent support for an
attractive $K^-$ optical potential would be given, with all its implications
for astrophysics and the physics of kaonic atoms. Furthermore, these
experiments would also help to address the question about the nature of the
$\lam$.

We have also briefly discussed the behavior of the $\lam$ at temperatures
relevant for relativistic heavy ion collisions. In a simplified model, we found
that the mass $\lam$ also increases, again giving rise to an
attractive optical potential for the $K^-$. Such an attractive mean field
would be very helpful in order to explain measured kaon spectra at the AGS
\cite{FKB93}. One might also wonder, to which extent the now possible decay
of the $\lam$ into $K^-$  and proton contributes to the recently observed very
cold component in the $K^-$ spectrum \cite{Sta94}. We believe that the
contribution is too small, because with a mass of the $\lam$  just above
threshold, the branching ratio into $K^-p$ is very small. In  addition, the
mass of the $\lam$ is fairly high, so that very few $\lam$ will be produced.
Thus, it seems very unlikely that enough $K^-$ will come from the $\lam$ to
account for the measured enhancement. Furthermore, if the reported slope
parameter of $\sim 15 \, \rm MeV$ is correct, the $\lam$ has to have a
temperature has low as $\sim 50 \, \rm MeV$, which is to be compared to the
canonical value of $\sim 150 \, \rm MeV$ seen in all other particle spectra.

The natural extension of this work is to include the mean field potential of
the nucleon and the $K^-$ in a selfconsistent
way along the lines of the Brueckner
Hartree Fock scheme and to carry out the full coupled channel problem also at
finite temperature. Such a calculation should be based on a realistic
interaction for the relevant channels such as e.g. the J\"ulich approach. A
detailed comparison with data from an experiment as proposed above will then
provide extremely interesting insight into the nature of the $\lam$ and the
properties of a $K^-$ in dense matter.

\noindent
{\bf Acknowledgements:}\\
I would like to thank G.F. Bertsch and G.E. Brown
for useful discussions. Parts of this work
have been carried out at the National Institute for Nuclear Theory in Seattle,
and its hospitality is greatly acknowledged. This work has been supported in
part by the US. Dept. of Energy Grant No. DE-FG02-88ER40388.

\newpage
\noindent
{\bf Figure captions}

\begin{figure}[h]
\caption{Comparison of the results for
the  forward $I=0$ scattering amplitude with the analysis
of Martin \protect\cite{Mar81}.}
\label{fig.2.1}
\end{figure}

\begin{figure}[h]
\caption{Mass of the $\lam$ as function of density for the one channel model
( eq. \protect\raf{eq.3.2}).}
\label{fig.3.1}
\end{figure}

\begin{figure}[h]
\caption{Forward $I=0$ scattering amplitude at finite densities for a cutoff of
$\Lambda = 1 \, \rm GeV$. The long dashed corresponds to the result at $\rho =
0.5 \rho_0$ for a cutoff of $\Lambda = 0.78 \, \rm GeV$. The arrow indicates
the $K^-N$ threshold.}
\label{fig.3.2}
\end{figure}

\begin{figure}[h]
\caption{Resulting optical potential for an s-wave $K^-$ in matter.
}
\label{fig.3.3}
\end{figure}

\begin{figure}[h]
\caption{$K^+$ missing mass spectrum for unshifted $\lam$ (full histogram) and
for a $\lam$ shifted in mass by $30 \, \rm MeV$ at nuclear matter density.
}
\label{fig.4.1}
\end{figure}

\end{document}